\documentclass[11pt,a4paper]{article}
\usepackage{amssymb,amsmath}
\numberwithin{equation}{section}
\newtheorem{definition}{Definition}[section]
\newtheorem{proposition}[definition]{Proposition}

\newtheorem{theorem}[definition]{Theorem}

\begin{document}
\title{Global existence and asymptotic behaviour in the future for
the Einstein-Vlasov system with positive cosmological constant}
\author{Sophonie Blaise Tchapnda N.$^1$ and Alan D. Rendall$^2$   \\
$^{1}$Department of Mathematics, Faculty of
Science, \\University of Yaounde I, PO Box 812, Yaounde, Cameroon \\
{tchapnda@uycdc.uninet.cm}\\
$^{2}$Max-Planck-Institut f\"ur
Gravitationsphysik \\ Am M\"uhlenberg 1, D-14476 Golm, Germany \\
{rendall@aei.mpg.de}}
\date{}
\maketitle
\begin{abstract}
The behaviour of expanding cosmological models with collisionless matter
and a positive cosmological constant is analysed. It is shown that
under the assumption of plane or hyperbolic symmetry the area radius 
goes to infinity, the spacetimes are future geodesically complete, 
and the expansion becomes isotropic and exponential at late times.
This proves a form of the cosmic no hair theorem in this class of
spacetimes.
\end{abstract}
\section{Introduction}
The presence of a positive cosmological constant $\Lambda$ can lead to 
exponential expansion in cosmological models. This is the simplest 
mathematical description of an inflationary universe. Under certain 
circumstances the de Sitter solution acts as a late time attractor 
for more general solutions of the Einstein equations with $\Lambda>0$.
This is sometimes known as the cosmic no hair theorem. Up to now there 
are unfortunately not many cases where this kind of statement has been 
proved rigorously for inhomogeneous spacetimes. 

A positive cosmological constant can be introduced in Newtonian cosmology
and this provides a simplified model for the general relativistic case.
In \cite{brauer} a form of the cosmic no hair theorem was proved in Newtonian
cosmology. A perfect fluid was used as a matter model and solutions were
considered which evolve from initial data which are small but finite
perturbations of homogeneous data. It was shown that if the homogeneous
solution exists globally in the future the same is true of the 
inhomogeneous solution. Of course a global in time existence theorem is
a prerequisite for a proof of the cosmic no hair theorem. It was then 
shown that the inhomogeneous solutions have a behaviour at late times which
is qualitatively similar to that of the homogeneous model. If $\bar\rho$
is the mean density and $\delta\rho=\rho-\bar\rho$ then $\delta\rho/\bar\rho$ 
converges as $t\to\infty$. In Newtonian cosmology there is also a theorem 
about the late time asymptotics of models with a kinetic description of 
matter by the Vlasov equation and with vanishing cosmological constant 
\cite{rein3}. The boundedness of $\delta\rho/\bar\rho$ is also obtained in
that case. Adding a positive cosmological constant to the problem 
considered in \cite{rein3} would presumably simplify the analysis but this 
has not been attempted.

In general relativity the problem of proving the cosmic no hair theorem 
is more difficult. In the spatially homogeneous case there is a general 
result of Wald \cite{wald} on spacetimes with positive cosmological constant 
which does not depend on the details of the matter content but only on energy
conditions. There is one example where future geodesic completeness has
been proved for a class of inhomogeneous spacetimes with matter \cite{rein2}.
This concerns solutions of the Einstein-Vlasov system with hyperbolic 
symmetry and $\Lambda=0$ satisfying an additional inequality on the initial 
data. Under those assumptions geodesic completeness was proved but only 
limited information was obtained on the asymptotic behaviour at late times. 
In the following we will show that in the presence of a positive cosmological 
constant this result can be strengthened a lot. Future geodesic completeness 
is proved for all solutions with hyperbolic and plane symmetry. Moreover the 
asymptotic behaviour is shown to closely resemble that of the de Sitter 
solution. It should be mentioned that in the case of the vacuum Einstein 
equations there is a proof of a form of the cosmic no hair theorem which 
does not require symmetry assumptions but does require a small data 
restriction \cite{friedrich}.

In this paper we study solutions of the Einstein equations with positive
cosmological constant coupled to the Vlasov equation describing collisionless 
matter. Under the assumption of plane or hyperbolic symmetry we show that
the solutions are future geodesically complete and we obtain a detailed 
description of their late time behaviour, which is similar to that of the
de Sitter solution. The proof is built on the local existence theorem and
continuation criterion previously proved by Tchapnda and Noutchegueme
\cite{tchapnda}.

Let us recall the formulation of the Einstein-Vlasov system which governs 
the time evolution of a self-gravitating collisionless gas in the context
of general relativity; for the moment we do not assume any
symmetry of the spacetime. All the particles in the gas are
assumed to have equal rest mass, normalized to unity. The four-momentum of
each particle is a future-pointing unit timelike vector so that the number 
density $f$ of particles is a non-negative function supported on the mass 
shell
\begin{align*}
PM:= \{g_{\alpha\beta}p^{\alpha}p^{\beta} = -1,
 \ p^0 > 0 \},
\end{align*}
a submanifold of tangent bundle $TM$ of the space-time manifold
$M$ with metric $g$ (the signature is $-+++$). We use coordinates
$(t,x^a)$ with zero shift and corresponding canonical momenta
$p^\alpha$ ; Greek indices always run from $0$ to $3$, and Latin
ones from $1$ to $3$. On the mass shell $PM$ the variable $p^0$
becomes a function of the remaining variables $(t, x^a, p^b)$ :
\begin{align*}
p^0 = \sqrt{-g^{00}}\sqrt{1+g_{ab}p^{a}p^{b}}.
\end{align*}
The Einstein-Vlasov system now reads
\begin{align*}
\partial_{t}f + \frac{p^{a}}{p^{0}} \partial_{x^{a}}f -
\frac{1}{p^{0}}\Gamma_{\beta\gamma}^{a} p^{\beta} p^{\gamma}
\partial_{p^{a}}f
= 0 \\ G_{\alpha\beta} + \Lambda g_{\alpha\beta} = 8 \pi T_{\alpha\beta} \\
T_{\alpha\beta} = - \int_{\mathbb{R}^{3}}f p_{\alpha}
p_{\beta}|g|^{1/2} \frac{dp^{1}dp^{2}dp^{3}}{p_{0}}
\end{align*}
where $p_{\alpha} = g_{\alpha\beta} p^{\beta}$,
$\Gamma_{\beta\gamma}^{\alpha}$ are the Christoffel symbols, $|g|$
denotes the determinant of the metric $g$, $G_{\alpha\beta}$ the
Einstein tensor, $\Lambda$ the cosmological constant, and
$T_{\alpha\beta}$ is the energy-momentum tensor.

In this paper we want to investigate the Einstein-Vlasov system
with  positive cosmological constant and plane or hyperbolic
symmetry in the expanding direction. For the notion of spherical,
plane or hyperbolic symmetry, we refer to \cite{rendall1}. We write
the system in areal coordinates, i.e., the coordinates are chosen
such that $R = t$, where $R$ is the area radius function on a
surface of symmetry. 

The circumstances under which coordinates of this type exist are discussed 
in \cite{andreasson2} for the Einstein-Vlasov system with vanishing
cosmological constant. It will now be shown that the analysis there can be
extended to the situation under consideration here. Consider first the
Einstein equations with a general matter model satisfying the dominant 
energy condition and $\Lambda=0$. For
plane symmetric spacetimes it follows from Proposition 3.1 of \cite{rendall2} 
that the gradient of $R$ is always timelike. The corresponding statment in 
the case of hyperbolic symmetry can be proved by the argument in Step 1 in
section 4 of \cite{andreasson2}. When there is a cosmological constant
we can consider it as a fictitious matter field with \lq energy-momentum
tensor\rq\ $-\Lambda g_{\alpha\beta}$. This fictitious matter field
satisfies the dominant energy condition. The same is true of the 
tensor which is the sum of the fictitious energy-momentum tensor with
the energy-momentum tensor of real matter satisfying the dominant energy
condition. It can be concluded from all this that the gradient of $R$
is timelike for the Einstein-Vlasov system with positive cosmological
constant and plane or hyperbolic symmetry. The remainder of Step 1 and 
Step 2 in section 4 of 
\cite{andreasson1} and \cite{andreasson2} can be extended to the case
where a positive cosmological constant is present using the same
method of the fictitious energy-momentum tensor. The only property
which is required in addition to the dominant energy condition is the
inequality $q\le\rho-p$ and this is satisfied by the fictitous 
energy-momentum tensor. From this point it is possible to argue exactly
as in \cite{andreasson1} and \cite{andreasson2} to conclude that a
solution of the Einstein-Vlasov system with positive cosmological constant
and plane or hyperbolic symmetry contains a Cauchy surface where $R$ is
constant. Hence there is no loss of generality in restricting 
consideration to spacetimes evolving from a hypersurface of constant areal 
time.

Since the gradient of $R$ is everywhere timelike it must be either
everywhere future-pointing timelike or everywhere past-pointing timelike.
We choose a time orientation such that the latter is the case. Then the
expanding direction of the cosmological model corresponding to increasing
area radius $t$. 
 
The metric takes the form
\begin{equation} \label{eq:1.1}
  ds^2 = -e^{2\mu(t,r)}dt^2 + e^{2\lambda(t,r)}dr^2 + t^2
  (d\theta^2 + \sin_{k}^{2}\theta d\varphi^{2})
\end{equation}
Here $t > 0$, the functions $\lambda$ and $\mu$ are periodic in
$r$ with period $1$ and
\begin{displaymath}
 \sin_{k}\theta := \left\{ \begin{array}{ll}
\sin\theta & \textrm{if $k=1$}\\
1 & \textrm{if $k=0$}\\
\sinh\theta & \textrm{if $k=-1$}
  \end{array} \right.
\end{displaymath}
For the case $k = 1$ the orbits of the symmetry action are
two-dimensional spheres. In this spherically symmetric case, as
well as in the case $\Lambda<0$, the global result below is seen
to be false, cf. \cite{tchapnda}, so these cases will not be
considered further. For the plane symmetric case $k=0$ the orbits
of the symmetry action are flat tori, they are hyperbolic spaces
for the hyperbolic symmetry $k=-1$. The coordinates $(\theta,
\varphi)$ range in $[0, 2\pi] \times [0, 2\pi]$ or $[0, \infty[
\times [0, 2\pi]$ for $k = 0$, $-1$ respectively. It has been
shown in \cite{rein1} and \cite{andreasson2} that due to the
symmetry $f$ can be written as a function of
\begin{align*}
t, r, w := e^{\lambda}p^1 \ & \textrm{and} \  F := t^{4}(p^2)^2 +
t^4 \sin_{k}^{2}\theta (p^{3})^{2}  ,
\end{align*}
i.e. $f = f(t, r, w, F)$. In these variables we have $p^0 =
e^{-\mu}\sqrt{1 + w^{2} + F/t^{2}}$. After calculating the Vlasov
equation in these variables, the non-trivial components of the
Einstein tensor, and the energy-momentum tensor and denoting by a
dot or by prime the derivatives of the metric components with
respect to $t$ or $r$ respectively, the complete Einstein-Vlasov
system reads as follows :
\begin{equation} \label{eq:1.2}
\partial_{t}f + \frac{e^{\mu-\lambda}w}{\sqrt{1+w^{2}+F/t^{2}}}
\partial_{r}f - (\dot{\lambda}w +
e^{\mu-\lambda}\mu'\sqrt{1+w^{2}+F/t^{2}})\partial_{w}f = 0
\end{equation}
\begin{equation} \label{eq:1.3}
e^{-2\mu} (2t\dot{\lambda}+1)+ k - \Lambda t^{2} = 8 \pi t^{2}\rho
\end{equation}
\begin{equation} \label{eq:1.4}
e^{-2\mu} (2t\dot{\mu}-1)- k + \Lambda t^{2} = 8 \pi t^{2}p
\end{equation}
\begin{equation} \label{eq:1.5}
\mu' = -4 \pi t e^{\lambda+\mu}j
\end{equation}
\begin{equation} \label{eq:1.6}
e^{-2\lambda}\left(\mu'' + \mu'(\mu' - \lambda')\right) -
e^{-2\mu}\left(\ddot{\lambda}+(\dot{\lambda}-
\dot{\mu})(\dot{\lambda}+\frac{1}{t})\right) + \Lambda  = 4 \pi q
\end{equation}
where
\begin{equation} \label{eq:1.7}
\rho(t, r) := \frac{\pi}{t^{2}} \int_{-\infty}^{\infty}
\int_{0}^{\infty} \sqrt{1+w^{2}+F/t^{2}} f(t, r, w, F) dF dw =
e^{-2\mu}T_{00}(t, r)
\end{equation}
\begin{equation} \label{eq:1.8}
p(t, r) := \frac{\pi}{t^{2}} \int_{-\infty}^{\infty}
\int_{0}^{\infty} \frac{w^{2}}{\sqrt{1+w^{2}+F/t^{2}}} f(t, r, w,
F) dF dw = e^{-2\lambda}T_{11}(t, r)
\end{equation}
\begin{equation} \label{eq:1.9}
j(t, r) := \frac{\pi}{t^{2}} \int_{-\infty}^{\infty}
\int_{0}^{\infty} w f(t, r, w, F) dF dw = -e^{\lambda +
\mu}T_{01}(t, r)
\end{equation}
\begin{equation} \label{eq:1.10}
q(t, r) := \frac{\pi}{t^{4}} \int_{-\infty}^{\infty}
\int_{0}^{\infty} \frac{F}{\sqrt{1+w^{2}+F/t^{2}}} f(t, r, w, F)
dF dw = \frac{2}{t^{2}}T_{22}(t, r).
\end{equation}
We prescribe initial data at some time $t = t_0 > 0$,
\begin{eqnarray*}
f(t_0, r, w, F)= \overset{\circ}{f}(r, w, F), \ \lambda(t_0, r) =
\overset{\circ}{\lambda}(r) , \ \mu(t_0, r) =
\overset{\circ}{\mu}(r).
\end{eqnarray*}

The organization of the paper is as follows. In the next section
we show that the solution of the initial value problem
corresponding to the Einstein-Vlasov system
(\ref{eq:1.2})-(\ref{eq:1.6}) exists for all $t \geq t_0$. In
section $3$ we first prove that the spacetime obtained in section
$2$ is timelike and null geodesically complete towards the future;
later on we determine  the explicit leading behaviour of
solution, compute the generalized Kasner exponents and prove that
each of them tends to $\frac{1}{3}$ as $t$ tends to $+\infty$.

\section{Global existence in the future}

In this section we follow the approach of \cite{andreasson2}. We
make use of the continuation criterion in the following local
existence result :
\begin{proposition} \label{p:2.1} Let
 $\overset{\circ}{f} \in C^{1}(\mathbb{R}^{2} \times [0, \infty[)$
 with $\overset{\circ}{f}(r+1,w,F) = \overset{\circ}{f}(r,w,F)$
 for $(r,w,F) \in \mathbb{R}^{2} \times [0, \infty[$,
 $\overset{\circ}{f}\geq 0$, and
 \begin{eqnarray*}
 w_0 := \sup \{ |w| | (r,w,F) \in {\rm supp} \overset{\circ}{f} \} <
 \infty
 \end{eqnarray*}
\begin{eqnarray*}
 F_0 := \sup \{ F | (r,w,F) \in {\rm supp} \overset{\circ}{f} \} <
 \infty
 \end{eqnarray*}
 Let $\overset{\circ}{\lambda} \in C^{1}(\mathbb{R})$,
 $\overset{\circ}{\mu} \in C^{2}(\mathbb{R})$ with
 $\overset{\circ}{\lambda}(r) = \overset{\circ}{\lambda}(r+1)$,
$\overset{\circ}{\mu}(r) = \overset{\circ}{\mu}(r+1)$ for $r \in
\mathbb{R}$, and
\begin{eqnarray*}
\overset{\circ}{\mu}'(r) =
 -4 \pi t_0 e^{\overset{\circ}{\lambda} +
\overset{\circ}{\mu}}\overset{\circ}{j}(r) = -\frac{4
\pi^{2}}{t_0}e^{\overset{\circ}{\lambda}+ \overset{\circ}{\mu}}
\int_{-\infty}^{\infty} \int_{0}^{\infty}w
\overset{\circ}{f}(r,w,F) dF dw, \ \  r \in \mathbb{R}.
\end{eqnarray*}
Then there exists a unique, right maximal, regular solution $(f,
\lambda, \mu)$ of (\ref{eq:1.2})-(\ref{eq:1.6}) with $(f, \lambda,
\mu)(t_0) = (\overset{\circ}{f}, \overset{\circ}{\lambda},
\overset{\circ}{\mu})$ on a time interval $[t_0, T[$ with $T \in ]
t_0, \infty]$. If
\begin{equation*}
\sup \left\{ \mu(t, r) | r \in \mathbb{R}, t \in [t_0, T[ \right\}
< \infty
\end{equation*}
then $T = \infty$.
\end{proposition}
This is the content of  Thms. 3.3 and 3.4 in \cite{tchapnda}. For
a {\it regular solution} all derivatives which appear in the
system exist and are continuous by definition, cf.
\cite{tchapnda}.

We now establish a series of estimates which will result in an
upper bound on $\mu$ and will therefore prove that $T = \infty$.
Similar estimates were used in \cite{andreasson1} for the
Einstein-Vlasov system with Gowdy symmetry and were generalized 
to the case of $T^2$ symmetry in \cite{andreasson3}. Unless otherwise
specified in what follows constants denoted by $C$ will be
positive, may depend on the initial data and on $\Lambda$ and may
change their value from line to line.

Firstly, integration of (\ref{eq:1.4}) with respect to $t$ and the
fact that $p$ is non-negative imply that
\begin{align} \label{eq:2.1}
e^{2 \mu(t, r)} & = \left[\frac{t_{0}(e^{-2
\overset{\circ}{\mu}(r)} + k)}{t} - k - \frac{8
\pi}{t}\int_{t_0}^{t}s^{2}p(s, r) ds + \frac{\Lambda}{3t}(t^{3} -
t_{0}^{3} )\right]^{-1} \nonumber\\
& \geq  \frac{t}{C-kt+\frac{\Lambda}{3}t^{3}}, \ t \in [t_0, T[.
\end{align}
In this inequality, $C$ does not depend on $\Lambda$. Next we
claim that
\begin{equation}\label{eq:2.2}
\int_{0}^{1} e^{\mu+\lambda}\rho(t,r) dr \leq C t^{-3}, \ t \in
[t_0, T[.
\end{equation}
A lengthy computation shows that
\begin{equation}\label{eq:2.3}
\frac{d}{dt}\int_{0}^{1} e^{\mu+\lambda}\rho(t,r) dr =
-\frac{1}{t}\int_{0}^{1} e^{\mu+\lambda}\left[2\rho + q
-\frac{\rho+p}{2}(1+ke^{2\mu}-\Lambda t^{2} e^{2\mu})\right] dr.
\end{equation}
Now using (\ref{eq:2.1}) we have
\begin{equation*}
1+ke^{2\mu}-\Lambda t^{2} e^{2\mu} \leq 1+ \frac{kt - \Lambda
t^{3}}{C - kt + \frac{\Lambda}{3} t^{3}} = \frac{C -
\frac{2}{3}\Lambda t^{3}}{C - kt + \frac{\Lambda}{3} t^{3}}.
\end{equation*}
The right hand side of this inequality is negative if $t \geq
(\frac{3C}{2\Lambda})^{1/3}$. In this case $1+ke^{2\mu}-\Lambda
t^{2} e^{2\mu} \leq 0$ so that, using the fact that $q \geq 0$ and
$p \geq 0$, (\ref{eq:2.3}) implies that
\begin{equation}\label{eq:2.4}
\frac{d}{dt}\int_{0}^{1} e^{\mu+\lambda}\rho(t,r) dr \leq
-\frac{1}{t}\int_{0}^{1} e^{\mu+\lambda}\left[2\rho -
\frac{\rho}{2}(1+ke^{2\mu}-\Lambda t^{2} e^{2\mu})\right] dr.
\end{equation}
Setting $C'(\Lambda):= \frac{3}{\Lambda}(3C-2k)$, we have the
estimate
\begin{equation*}
1+ke^{2\mu}-\Lambda t^{2} e^{2\mu} \leq 1+ \frac{kt - \Lambda
t^{3}}{C - kt + \frac{\Lambda}{3} t^{3}} \leq C'(\Lambda)t^{-2}-2,
\end{equation*}
and combining this with (\ref{eq:2.4}) yields
\begin{equation*}
\frac{d}{dt}\int_{0}^{1} e^{\mu+\lambda}\rho(t,r) dr \leq
-\frac{3}{t}\int_{0}^{1} e^{\mu+\lambda}\rho dr +
\frac{C'(\Lambda)}{2t^{3}} \int_{0}^{1} e^{\mu+\lambda}\rho(t,r)
dr
\end{equation*}
which multiplied by $t^{3}$ gives
\begin{equation*}
\frac{d}{dt}\left[t^{3}\int_{0}^{1} e^{\mu+\lambda}\rho dr\right]
\leq Ct^{-3}\left[t^{3}\int_{0}^{1} e^{\mu+\lambda}\rho dr\right]
\end{equation*}
By Gronwall's inequality, this implies that
\begin{equation*}
\int_{0}^{1} e^{\mu+\lambda}\rho(t,r) dr \leq Ct^{-3},
\end{equation*}
i.e. (\ref{eq:2.2}) for $t \geq (\frac{3C}{2\Lambda})^{1/3}$.
For $t < (\frac{3C}{2\Lambda})^{1/3}$, (\ref{eq:2.3}) implies the
following, since $q \geq 0$ :
\begin{align*}
\frac{d}{dt}\int_{0}^{1} e^{\mu+\lambda}\rho(t,r) dr & \leq
-\frac{2}{t}\int_{0}^{1} e^{\mu+\lambda}\rho dr +
\frac{1}{t}\int_{0}^{1}e^{\mu+\lambda}\frac{\rho+p}{2}dr
\nonumber\\
& {} + \frac{1}{t}\int_{0}^{1}\frac{(k-\Lambda
t^{2})e^{2\mu}e^{\mu+\lambda}}{2}(\rho +p)dr \nonumber\\
& \leq -\frac{1}{t}\int_{0}^{1} e^{\mu+\lambda}\rho dr,
\end{align*}
we have used the fact that $\rho \geq p$ and $k-\Lambda t^{2} \leq
0$. By Gronwall's inequality, we obtain
\begin{align*}
\int_{0}^{1} e^{\mu+\lambda}\rho(t,r) dr & \leq Ct^{-1} \\
& {} \leq (Ct^{2}+C)t^{-3}\\
& {} \leq \left[C(\frac{3C}{2\Lambda})^{2/3}+C\right]t^{-3} \
\textrm{since} \ t<(\frac{3C}{2\Lambda})^{1/3}
\end{align*}
that is (\ref{eq:2.2}) holds for $t<(\frac{3C}{2\Lambda})^{1/3}$
as well. Using (\ref{eq:2.2}) and the equation $\mu' = -4\pi
te^{\mu+\lambda}j$ we find
\begin{align}\label{eq:2.5}
\mid \mu(t,r)- \int_{0}^{1}\mu(t,\sigma)d\sigma \mid & =
 \mid \int_{0}^{1}\int_{\sigma}^{r}\mu'(t,\tau)d\tau d\sigma \mid
 \leq \int_{0}^{1}\int_{0}^{1}|\mu'(t,\tau)|d\tau d\sigma
 \nonumber\\
 & \leq 4\pi t\int_{0}^{1}e^{\mu+\lambda}|j(t,\tau)|d\tau
 \leq 4\pi t\int_{0}^{1}e^{\mu+\lambda}\rho(t,\tau)d\tau
 \nonumber\\
 & \leq Ct^{-2}, \ t \in [t_0, T[, \ r \in [0,1].
\end{align}
Next we show that
\begin{equation}\label{eq:2.6}
  e^{\mu(t,r)-\lambda(t,r)} \leq Ct^{-2}, \ t \in [t_0, T[, \ r
  \in [0,1]
\end{equation}
To see this observe that by (\ref{eq:1.3}), (\ref{eq:1.4}) and
(\ref{eq:2.1})
\begin{align*}
 \frac{ \partial}{\partial t}e^{\mu-\lambda} & =
 e^{\mu-\lambda}\left[4\pi te^{2\mu}(p-\rho)+\frac{1+ke^{2\mu}}{t}-\Lambda t
 e^{2\mu}\right]\leq e^{\mu-\lambda}\left[\frac{1+ke^{2\mu}}{t}-\Lambda t
 e^{2\mu}\right]\\
 & \leq \left[ \frac{1}{t} + \frac{k-\Lambda
 t^{2}}{C-kt+\frac{\Lambda}{3}t^{3}}\right]e^{\mu-\lambda}.
\end{align*}
Using the fact that $-k+\Lambda t^{2}$ is the derivative of
$C-kt+\frac{\Lambda}{3}t^{3}$ and integrating this inequality with
respect to $t$ yields
\begin{equation*}
e^{\mu-\lambda} \leq C \frac{t}{C-kt+\frac{\Lambda}{3}t^{3}} \leq
Ct^{-2},
\end{equation*}
i.e. (\ref{eq:2.6}).

We now estimate the average of $\mu$ over the interval $[0,1]$
which in combination with (\ref{eq:2.5}) will yield the desired
upper bound on $\mu$ :
\begin{align*}
\int_{0}^{1}\mu(t,r) dr & = \int_{0}^{1}\overset{\circ}{\mu}(r) dr
+ \int_{t_0}^{t}\int_{0}^{1}\dot{\mu}(s,r) dr ds\\
& \leq C+\int_{t_0}^{t}\frac{1}{2s}\int_{0}^{1}[e^{2\mu}(8\pi
s^{2}p + k-\Lambda s^{2})+1]dr ds\\
& \leq
C+\frac{1}{2}\ln(t/t_{0})+C\int_{t_0}^{t}s^{-4}ds
-\frac{1}{2}\int_{t_0}^{t}\frac{-k+\Lambda
s^{2} }{C-ks+\frac{\Lambda}{3}s^{3}}ds\\
& \leq C + \frac{1}{2}\left[\ln
\frac{s}{C-ks+\frac{\Lambda}{3}s^{3}}\right]_{s=t_0}^{s=t}
\end{align*}
where we used (\ref{eq:2.1}), (\ref{eq:2.2}), (\ref{eq:2.6}) and
the fact that $p \leq \rho$. With (\ref{eq:2.5}) this implies
\begin{equation}\label{eq:2.7}
\mu(t,r) \leq C(1+t^{-2}+\ln t^{-2}) \leq C, \ t \in [t_0, T[, \ r
\in [0,1]
\end{equation}
which by Proposition \ref{p:2.1} implies $T = \infty$. Thus we
have proven :
\begin{theorem} \label{t:2.2} For initial data as in Proposition
\ref{p:2.1} the solution of the Einstein-Vlasov system with
positive cosmological constant and plane or hyperbolic symmetry,
written in areal coordinates, exists for all $t \in [t_0, \infty[$
where $t$ denotes the area radius of the surfaces of symmetry of
the induced spacetime. The solution satisfies the estimates
(\ref{eq:2.2}), (\ref{eq:2.6}) and (\ref{eq:2.7}).
\end{theorem}

\section{On future asymptotic behaviour}

In the first part of this section we prove that the spacetime
obtained in Theorem \ref{t:2.2} is timelike and null geodesically
complete in the expanding direction. The analogue of this result
was proved by Rein, cf. \cite{rein2}, in the case $\Lambda = 0$,
$k=-1$ but with initial data satisfying a certain size
restriction, an additional assumption which we drop here due to
the fact that $\Lambda$ does not vanish. The proofs of the results
obtained in the first two subsections are modelled on the approach
of \cite{rein2}. In this section we are interested in proving
statements about the asymptotic behaviour of solutions at late
times. Therefore there is no loss of generality in prescribing data
at some large time $t = t_0 >0$.

Firstly we establish a bound on $w$ along characteristics of the
Vlasov equation.

\subsection{An estimate along characteristics}

Let
\begin{eqnarray*}
  w_0 := \sup \{ |w| | (r,w,F) \in {\rm supp} \overset{\circ}{f} \} <
 \infty,
 \end{eqnarray*}
\begin{eqnarray*}
  F_0 := \sup \{ F | (r,w,F) \in {\rm supp} \overset{\circ}{f} \} <
 \infty.
 \end{eqnarray*}
Except in the vacuum case we have $w_0 >0$ and $F_0 >0$. For $ t \geq
t_0 $ define
\begin{eqnarray*}
 P_{+}(t) := \max \{0, \max \{w | (r,w,F) \in {\rm supp} f(t) \} \},
\end{eqnarray*}
\begin{eqnarray*}
 P_{-}(t) := \min \{0, \min \{ w | (r,w,F) \in {\rm supp} f(t) \}
 \}.
\end{eqnarray*}
Fix $\varepsilon \in ]0,1[$. We claim that
\begin{equation}\label{eq:3.1}
  P_{+}(t) \leq w_{0}\left(\frac{t}{t_0}\right)^{-1+\varepsilon}, \
  P_{-}(t) \geq -w_{0}\left(\frac{t}{t_0}\right)^{-1+\varepsilon}, \ t \geq
  t_0.
\end{equation}
Assume that the estimate on $P_{+}$ were false for some $t$.
Define
\begin{equation*}
t_1 := \sup \left\{ t \geq t_0 | P_{+}(s) \leq
w_{0}\left(\frac{s}{t_0}\right)^{-1+\varepsilon}, t_0 \leq s \leq
t \right\}
\end{equation*}
so that $t_0 \leq t_1 < \infty$ and $P_{+}(t_1) =
w_{0}\left(\frac{t_1}{t_0}\right)^{-1+\varepsilon} > 0$. Choose
$\alpha \in ]0,1[$. By continuity, there exists some $t_2 > t_1$
such that the following holds :
\begin{equation*}
  (1-\alpha) P_{+}(s) > 0, \ s \in [t_1, t_2].
\end{equation*}
If for some characteristic curve $(r(s), w(s), F)$ in the
support of $f$, that is with $(r(t_0), w(t_0), F) \in {\rm
supp}\overset{\circ}{f}$, and for some $t \in ]t_1, t_2]$ the
estimate
\begin{equation}\label{eq:3.2}
(1-\alpha/2) P_{+}(t) \leq w(t) \leq P_{+}(t)
\end{equation}
holds then
\begin{equation}\label{eq:3.3}
(1-\alpha) P_{+}(s) \leq w(s) \leq P_{+}(s), \ s \in [t_1, t].
\end{equation}
Note that the estimates on $w$ from above hold by definition of
$P_+$ in any case. Let $(r(s), w(s), F)$ be a characteristic in
the support of $f$ satisfying (\ref{eq:3.2}) for some $t \in ]t_1,
t_2]$ and thus (\ref{eq:3.3}) on $[t_1, t]$. Then on $[t_1, t]$,
\begin{align}\label{eq:3.4}
\dot{w} & = \frac{4
\pi^{2}}{s}e^{2\mu}\int_{-\infty}^{\infty}\int_{0}^{\infty}
\left(\tilde{w}\sqrt{1+w^{2}+F/s^{2}}-w\sqrt{1+\tilde{w}^{2}+\tilde{F}/s^{2}}\right)f
d\tilde{F}d\tilde{w} \nonumber\\
& +\frac{1+ke^{2\mu}}{2s}w
-\frac{\Lambda}{2}swe^{2\mu} \nonumber\\
& \leq
\frac{4\pi^{2}}{s}e^{2\mu}\int_{0}^{P_{+}(s)}\int_{0}^{F_{0}}\frac{\tilde{w}^{2}
(1+w^{2}+F/s^{2})-w^{2}(1+\tilde{w}^{2}+\tilde{F}/s^{2})}{\tilde{w}
\sqrt{1+w^{2}+F/s^{2}}+w\sqrt{1+\tilde{w}^{2}+\tilde{F}/s^{2}}}f
d\tilde{F}d\tilde{w} \nonumber\\
& +\frac{1+ke^{2\mu}}{2s}w
-\frac{\Lambda}{2}swe^{2\mu} \nonumber\\
& \leq
\frac{4\pi^{2}}{s}e^{2\mu}\int_{0}^{P_{+}(s)}\int_{0}^{F_{0}}\frac{\tilde{w}(1+F)}{w}f
d\tilde{F}d\tilde{w}+\frac{1+ke^{2\mu}}{2s}w
-\frac{\Lambda}{2}swe^{2\mu} \nonumber\\
& \leq 4 \pi^{2}F_0 (1+F_0)\parallel
\overset{\circ}{f}\parallel\frac{e^{2\mu}}{2s}P_{+}^{2}(s)\frac{1}{w}+
\frac{1+ke^{2\mu}-\Lambda s^{2}e^{2\mu}}{2s}w \\
& \leq 4 \pi^{2}F_0 (1+F_0)\parallel
\overset{\circ}{f}\parallel\frac{1}{(1-\alpha)^{2}}\frac{e^{2\mu}}{2s}w
+ \frac{1+ke^{2\mu}-\Lambda s^{2}e^{2\mu}}{2s}w, \ \textrm{using
(\ref{eq:3.3})} \nonumber\\
& \leq \frac{1+(C+k-\Lambda s^{2})e^{2\mu}}{2s}w \nonumber .
\end{align}
Since $s$ is large, $C+k-\Lambda s^{2}$ is negative so that using
(\ref{eq:2.1}) we have
\begin{equation}\label{eq:3.5}
  \dot{w} \leq \frac{1+\frac{Cs+ks-\Lambda
  s^{3}}{C-ks+\frac{\Lambda}{3}s^{3}}}{2s}w.
\end{equation}
Now
\begin{align*}
 3+\frac{Cs+ks-\Lambda s^{3}}{C-ks+\frac{\Lambda}{3}s^{3}} & =
  \frac{3C+Cs-2ks}{C-ks+\frac{\Lambda}{3}s^{3}} \\
  & \leq \left(
  \frac{9C}{\Lambda}s^{-1}+\frac{3C}{\Lambda}-\frac{6k}{\Lambda}\right)s^{-2}
 \\
  & \leq \frac{C-6k}{\Lambda}s^{-2}
\end{align*}
so that setting $C'(\Lambda) := \frac{C-6k}{\Lambda}$ we obtain
the estimate
\begin{equation}\label{eq:3.6}
1+\frac{Cs+ks-\Lambda s^{3}}{C-ks+\frac{\Lambda}{3}s^{3}} \leq
C'(\Lambda)s^{-2}-2.
\end{equation}
Thus (\ref{eq:3.5}) implies that
\begin{equation*}
\dot{w} \leq -\frac{w}{s}+\frac{C'(\Lambda)}{2}s^{-3}w
\end{equation*}
which multiplied by $s^{1-\varepsilon}$ gives
\begin{align*}
\frac{d}{ds}(s^{1-\varepsilon}w) & \leq s^{1-\varepsilon}w
\left(-\varepsilon s^{-1} + \frac{C'(\Lambda)}{2}s^{-3}\right) \\
& \leq 0 \ \ \textrm{since} \ s \ \textrm{is large}.
\end{align*}
Thus the function $s \mapsto s^{1-\varepsilon}w(s)$ is decreasing
on $[t_1, t]$. This implies that
\begin{equation*}
  t^{1-\varepsilon}w(t) \leq t_{1}^{1-\varepsilon}w(t_{1}) \leq
  t_{1}^{1-\varepsilon}P_{+}(t_1) = \frac{w_0}{t_{0}^{-1+\varepsilon}}
\end{equation*}
by assumption on $t_1$ and so
\begin{equation}\label{eq:3.7}
  w(t) \leq w_0 \left( \frac{t}{t_0}\right)^{-1+\varepsilon}.
\end{equation}
This estimate holds only for characteristics which satisfy
(\ref{eq:3.2}), but this is sufficient to conclude that
\begin{equation*}
  P_{+}(t) \leq w_0 \left( \frac{t}{t_0}\right)^{-1+\varepsilon}, \ t
  \in [t_1, t_2],
\end{equation*}
in contradiction to the choice of $t_1$. The estimate on $P_{+}$
is now established. The analogous arguments for characteristics
with $w <0$ yield the assertion for $P_{-}$.\\ Next we consider
characteristics which are not in the support of $f$. We can
rewrite the inequality (\ref{eq:3.4}) for $s \in [t_0, t]$ and
$w(s) > 0$ :
\begin{equation*}
\dot{w} \leq 4 \pi^{2}F_0 (1+F_0)\parallel
\overset{\circ}{f}\parallel\frac{e^{2\mu}}{2s}P_{+}^{2}(s)\frac{1}{w}+
\frac{1+(k-\Lambda s^{2})e^{2\mu}}{2s}w.
\end{equation*}
From (\ref{eq:2.1}) and (\ref{eq:3.6}) it follows that
$\frac{1+(k-\Lambda s^{2})e^{2\mu}}{2s} \leq 0$. Using the estimate 
(\ref{eq:3.1}) on $P_{+}$ and (\ref{eq:2.7}) we obtain
\begin{equation*}
  \dot{w} \leq C s^{2\varepsilon-3}\frac{1}{2w}.
\end{equation*}
Hence
\begin{equation*}
  \frac{d}{ds}(w^{2}) \leq C s^{2\varepsilon-3}.
\end{equation*}
Integrating this over $[t_0, t]$ yields
\begin{equation}\label{eq:3.8}
  w^{2}(t) \leq C , \ t \geq t_0.
\end{equation}
The analogous arguments for characteristics outside the support of
$f$ with $w<0$ yield the same estimate. Thus by (\ref{eq:3.7}),
(\ref{eq:3.1}) and (\ref{eq:3.8}) we can state :
\begin{proposition} \label{p:3.1} For any characteristic $(r, w,
F)$, for any solution of Einstein-Vlasov system with positive
cosmological constant and plane or hyperbolic symmetry written in
areal coordinates and with initial data as in Proposition
\ref{p:2.1},
\begin{equation*}
  |w(t)| \leq C, \ t \geq t_0,
\end{equation*}
where the positive constant $C$ depends on the initial data.
\end{proposition}

\subsection{Geodesic completeness}
Let $]\tau_{-}, \tau_{+}[ \ni \tau \mapsto (x^{\alpha}(\tau),
p^{\beta}(\tau))$ be a geodesic whose existence interval is
maximally extended and such that $x^{0}(\tau_{0}) = t(\tau_{0}) =
t_0 $ for some $\tau_{0} \in ]\tau_{-}, \tau_{+}[$. We want to
show that for future-directed timelike and null geodesics 
$\tau_{+} = + \infty$. Consider first the case of a timelike
geodesic, i.e.,
\begin{equation*}
g_{\alpha\beta}p^{\alpha}p^{\beta} = -m^{2} \ ; \ p^{0}>0
\end{equation*}
with $m>0$. Since $\frac{dt}{d\tau} = p^{0}>0$, the geodesic can
be parametrized by the coordinate time $t$. With respect to
coordinate time the geodesic exists on the interval $[t_0,
\infty[$ since on bounded $t$-intervals the Christoffel symbols
are bounded and the right hand sides of the geodesic equations
written in coordinate time are linearly bounded in $p^{1}$, $
p^{2}$, $p^{3}$. Recall that along geodesics the variables $t$,
$r$, $p^{0}$, $w:=e^{\lambda}p^{1}$,
$F:=t^{4}\left[(p^{2})^{2}+\sin_{k}^{2}\theta(p^{3})^{2}\right]$
satisfy the following system of differential equations :
\begin{equation}\label{eq:3.9}
\frac{dr}{d\tau} = e^{-\lambda}w, \ \frac{dw}{d\tau} =
-\dot{\lambda}p^{0}w - e^{2\mu-\lambda}\mu'(p^{0})^{2}, \
\frac{dF}{d\tau} = 0
\end{equation}
\begin{equation}\label{eq:3.10}
\frac{dt}{d\tau} = p^{0}, \ \frac{dp^{0}}{d\tau} =
-\dot{\mu}(p^{0})^{2} -
2e^{-\lambda}\mu'p^{0}w-e^{-2\mu}\dot{\lambda}w^{2}-e^{-2\mu}t^{-3}F.
\end{equation}
Along the geodesic we define $w$ and $F$ as above. Then the
relation between coordinate time and proper time along the
geodesic is given by
\begin{equation*}
\frac{dt}{d\tau} = p^{0} = e^{-\mu}\sqrt{m^{2}+w^{2}+F/t^{2}},
\end{equation*}
and to control this we need to control $w$ as a function of
coordinate time. By (\ref{eq:2.1}) we have the estimate
\begin{equation*}
  e^{\mu} \geq Ct^{-1}, \ t\geq t_0.
\end{equation*}
Combining this with the estimate on $w$ in Proposition \ref{p:3.1}
yields the following along the geodesic :
\begin{equation*}
\frac{d\tau}{dt} = \frac{e^{\mu}}{\sqrt{m^{2}+w^{2}+F/t^{2}}} \geq
\frac{C t^{-1}}{\sqrt{m^{2}+C+F}}.
\end{equation*}
Since the integral of the right hand side over $[t_0, \infty[$
diverges, $\tau_{+} = + \infty$ as desired. In the case of a future-directed
null geodesic, i.e. $m=0$ and $p^{0}(\tau_{0})>0$, $p^0$ is everywhere
positive and the quantity $F$ is again conserved. The argument can now be 
carried out exactly as in the timelike case, implying that 
$\tau_{+}= +\infty$. We have proven :
\begin{theorem}\label{t:3.2} Consider initial data with plane or 
hyperbolic symmetry for the Einstein-Vlasov system with positive 
cosmological constant. Suppose that the regularity properties
required in the statement of Proposition \ref{p:2.1} are satisfied.
If the gradient of $R$ is initially past-pointing then there is a
corresponding Cauchy development which is future geodesically complete. 
\end{theorem}

\subsection{Determination of the leading asymptotic behaviour}
In this subsection we determine the explicit leading behaviour of
$\lambda$, $\mu$, $\dot{\lambda}$, $\dot{\mu}$, $\mu'$, and later
on we compute the generalized Kasner exponents and prove that each
of them tends to $1/3$ as $t$ tends to $+\infty$.

Let us recall briefly some of the relevant notation. Let $I$ be a
set of real numbers and $t_{*}$ a real number or infinity. The
asymptotic behaviour of a function $g$ defined on $I$ as $t
\rightarrow t_{*}$ is to be described. It will be compared with a
positive function $h(t)$, typically a power of $t$. The notation
$g(t) = O(h(t))$ as $t \rightarrow t_{*}$ means that there is a
neighbourhood $U$ of $t_*$ such that there is a constant $C$ with
$|g(t)| \leq Ch(t)$ for all $t$ belonging to both $I$ and $U$. The
notation $g(t) = o(h(t))$ as $t\rightarrow t_*$ means that
$g(t)/h(t)$ tends to $0$ as $t\rightarrow t_*$.

Now (\ref{eq:1.4}) can be written in the form
\begin{equation}\label{eq:3.11}
  \frac{d}{dt}(te^{-2\mu}) = \Lambda t^{2}-k-8\pi t^{2}p.
\end{equation}
Integrating this over $[t_0, t]$ yields
\begin{equation}\label{eq:3.12}
  te^{-2\mu}= (t_0 e^{-2\mu(t_0)}+k t_0 -
  \frac{\Lambda}{3}t_{0}^{3}) +
  \frac{\Lambda}{3}t^{3}-kt-\int_{t_0}^{t}8\pi s^{2}p ds.
\end{equation}
By (\ref{eq:3.7}) we have the following, where $C$ is a positive
constant and $\varepsilon \in ]0,1[$ :
\begin{equation*}
w \leq Ct^{-1+\varepsilon} \ \textrm{for} \ t \geq t_0.
\end{equation*}
Using the expression (\ref{eq:1.8}) for $p$, this implies that
\begin{equation*}
p \leq Ct^{-5+3\varepsilon}
\end{equation*}
so that
\begin{equation}\label{eq:3.13}
8\pi t^{2}p \leq Ct^{-3+3\varepsilon}.
\end{equation}
Assuming $\varepsilon<2/3$ we obtain, using (\ref{eq:3.12}) 
\begin{equation*}
  |te^{-2\mu}-\frac{\Lambda}{3}t^{3}+kt| \leq C,
\end{equation*}
i.e.,
\begin{equation*}
  e^{-2\mu} = \frac{\Lambda}{3}t^{2}\left(1+O(t^{-2})\right).
\end{equation*}
It follows that
\begin{equation}\label{eq:3.14}
  e^{\mu} =
  \sqrt{\frac{3}{\Lambda}}t^{-1}\left(1+O(t^{-2})\right).
\end{equation}
Now by (\ref{eq:1.3}), and using (\ref{eq:3.14}) and the fact that
$8\pi t \rho = O(t^{-2+\varepsilon})$, we have
\begin{align*}
\dot{\lambda} & = \frac{1}{2}(\Lambda t + 8\pi t
\rho)e^{2\mu}-\frac{1+ke^{2\mu}}{2t}\\
& = \frac{3}{2\Lambda}t^{-2}\left(1+O(t^{-2})\right)\left(\Lambda
t + O(t^{-2+\varepsilon})
\right)-\frac{1}{2t}-\frac{k}{2t}\left[\frac{3}{\Lambda}t^{-2}\left(1+O(t^{-2})\right)\right]
\end{align*}
and hence
\begin{equation}\label{eq:3.15}
  \dot{\lambda} = t^{-1}\left(1+O(t^{-2})\right).
\end{equation}
Integrating this over $[t_0, t]$ yields
\begin{equation}\label{eq:3.16}
 \lambda = \ln t \left[1+O\left((\ln t)^{-1}\right)\right].
\end{equation}
Next, using (\ref{eq:1.4}), (\ref{eq:3.13}) and (\ref{eq:3.14}) we
have
\begin{equation}\label{eq:3.17}
  \dot{\mu} = -t^{-1}\left(1+O(t^{-2})\right)
\end{equation}
and integrating this over $[t_0, t]$ yields
\begin{equation}\label{eq:3.18}
\mu = -\ln t\left[1+O\left((\ln t)^{-1}\right)\right].
\end{equation}
Now (\ref{eq:3.16}) implies that
\begin{equation*}
 e^{\lambda} = O(t),
\end{equation*}
the expression (\ref{eq:1.9}) of $j$ implies that
\begin{equation*}
|j| \leq C t^{-4+2\varepsilon}
\end{equation*}
and thus using equation (\ref{eq:1.5}) we obtain
\begin{equation}\label{eq:3.19}
\mu' = O(t^{-3+2\varepsilon}).
\end{equation}
We can now compute the limiting values of the generalized Kasner
exponents namely
\begin{equation*}
\frac{K_{1}^{1}(t,r)}{K(t,r)} = \frac{t \dot{\lambda}(t,r)}{t
\dot{\lambda}(t,r)+2}, \ \frac{K_{2}^{2}(t,r)}{K(t,r)} =
\frac{K_{3}^{3}(t,r)}{K(t,r)} = \frac{1}{t \dot{\lambda}(t,r)+2},
\end{equation*}
where $K(t,r) = K_{i}^{i}(t,r)$ is the trace of the second
fundamental form $K_{ij}$ of the metric. We refer to \cite{rein1}
for the computation. Using (\ref{eq:3.15}), we see that as $t$
tends to $+\infty$, each of those quantities tends to $1/3$
uniformly in $r \in \mathbb{R}$. We have proved the following :
\begin{theorem}\label{t:3.3} Let $(f,\lambda, \mu)$ be a solution
of the Einstein-Vlasov system with plane or hyperbolic symmetry
and $\Lambda >0$ given in the expanding direction. Then the
following properties hold at late times : (\ref{eq:3.15}),
(\ref{eq:3.16}),  (\ref{eq:3.17}),  (\ref{eq:3.18}),
(\ref{eq:3.19}), with $\varepsilon \in ]0,2/3[$ ; and
\begin{displaymath}
\lim_{t\rightarrow + \infty} \frac{K_{1}^{1}(t,r)}{K(t,r)} =
\lim_{t\rightarrow + \infty} \frac{K_{2}^{2}(t,r)}{K(t,r)} =
\lim_{t\rightarrow + \infty} \frac{K_{3}^{3}(t,r)}{K(t,r)} =
\frac{1}{3}.
\end{displaymath}
\end{theorem}

This theorem shows how the de Sitter solution acts as a model for the
dynamics of the class of solutions considered in this paper. For if
we set $\lambda=\ln t$, $\mu=-\ln t$ and $k=0$ the spacetime obtained
is the de Sitter solution. Thus the leading terms in the asymptotic 
expansions of the metric components are exactly the quantities defined
by the de Sitter spacetime.

\vskip 10pt\noindent \textbf{Acknowledgements} : The authors
acknowledge support by a research grant from the VolkswagenStiftung, 
Federal Republic of Germany. The major part of this work was carried out
while SBTN was enjoying the hospitality of the Max Planck
Institute for Gravitational Physics, Golm.

\end{document}